\documentclass[preprint,preprintnumbers,amsmath,amssymb]{revtex4}

\usepackage{graphicx,color}
\usepackage{bm}

\begin{document}

\newcommand{\OLD}[1]{{\tiny {\bf old:} #1 }}
\newcommand{\NEW}[1]{{ \it #1 }}
\renewcommand{\vec}[1]{\boldsymbol{#1}}
\newcommand{\w}{\omega}
\newcommand{\zrzn}{ZrZn$_2$}
\newcommand{\ZrZn}{ZrZn$_2$}
\newcommand{\uge}{UGe$_2$}
\newcommand{\MnSi}{MnSi\,-\,110 }
\newcommand{\Mnsi}{MnSi\,-\,111 }
\newcommand{\Tc}{$T_{c}$ }
\newcommand{\rhoxx}{$\rho_{xx}$ }
\newcommand{\rhoxy}{$\rho_{xy}$ }

\renewcommand{\floatpagefraction}{0.5}
\bibliographystyle{nature}




\title{Emergent electrodynamics of skyrmions in a chiral magnet}



\author{T. Schulz}
\affiliation{Physik Department E21, Technische Universit\"at M\"unchen, D-85748 Garching, Germany}

\author{R. Ritz}
\affiliation{Physik Department E21, Technische Universit\"at M\"unchen, D-85748 Garching, Germany}

\author{A. Bauer}
\affiliation{Physik Department E21, Technische Universit\"at M\"unchen, D-85748 Garching, Germany}

\author{M. Halder}
\affiliation{Physik Department E21, Technische Universit\"at M\"unchen, D-85748 Garching, Germany}

\author{M. Wagner}
\affiliation{Physik Department E21, Technische Universit\"at M\"unchen, D-85748 Garching, Germany}

\author{C. Franz}
\affiliation{Physik Department E21, Technische Universit\"at M\"unchen, D-85748 Garching, Germany}

\author{C. Pfleiderer}
\affiliation{Physik Department E21, Technische Universit\"at M\"unchen, D-85748 Garching, Germany}

\author{K. Everschor}
\affiliation{Institute for Theoretical Physics, Universit\"at zu K\"oln, D-50937 K\"oln, Germany}

\author{M. Garst}
\affiliation{Institute for Theoretical Physics, Universit\"at zu K\"oln, D-50937 K\"oln, Germany}

\author{A. Rosch}
\affiliation{Institute for Theoretical Physics, Universit\"at zu K\"oln, D-50937 K\"oln, Germany}

\date{\today}

\maketitle

\textbf{When an electron moves in a smoothly varying non-collinear magnetic structure, its spin-orientation adapts constantly, thereby inducing forces that act on both the magnetic structure and the electron. These forces may be described by electric and magnetic fields of an emergent electrodynamics \cite{volovik87,Yang09,Hai09,Barnes07}. The topologically quantized winding number of so-called skyrmions, i.e., certain magnetic whirls, discovered recently in chiral magnets \cite{mueh09,neub09,yu:2010} are theoretically predicted to induce exactly one quantum of emergent magnetic flux per skyrmion. A moving skyrmion is therefore expected to induce an emergent electric field following Faraday's law of induction, which inherits this topological quantization \cite{zang11}. Here we report Hall effect measurements, which establish quantitatively the predicted emergent electrodynamics. This allows to obtain quantitative evidence of the depinning of skyrmions from impurities at ultra-low current densities of only ${\rm10^6\,Am^{-2}}$ and their subsequent motion. The combination of exceptionally small current densities and simple transport measurements offers fundamental insights into the connection between emergent and real electrodynamics of skyrmions in chiral magnets, and promises to be important for applications in the long-term.} 

Skyrmion lattice phases (SLPs) in chiral magnets such as MnSi and other B20 transition metal compounds are a new form of magnetic order, composed of topologically protected vortex lines  (the skyrmions) with a non-zero winding number that are stabilized parallel to a small applied magnetic field. Skyrmion lattices in magnetic materials were discovered only recently by means of small angle neutron scattering (SANS)  \cite{mueh09,Adams11}. The winding number of the spin structure was thereby first inferred from topological contributions to the Hall effect \cite{neub09} and Lorentz force microscopy for thin samples, where the latter even established the existence of individual skyrmions \cite{yu:2010}. So far SLPs have been identified in all non-centrosymmetric B20 transition metal compounds that order helimagnetically at zero magnetic field regardless whether they are pure metals, strongly doped semiconductors  or even insulators \cite{mueh09,muen09,pfleiderer:JPCM2010}. The excellent theoretical understanding implies that skyrmions are a general phenomenon to be expected in a wide range of bulk materials as well as nano-scale systems \cite{bogd89,mueh09,yu:2010}, with the identification of spontaneous skyrmion lattices in monatomic layers of Fe on an Ir substrate as a major new development \cite{heinze11}.

Exploratory SANS studies  \cite{Jonietz10} revealed a rotation of the diffraction pattern in the SLP of single-crystal MnSi when an electric current applied transverse to the skyrmions exceeded an ultralow threshold of $j_c \sim10^{6}\,{\rm Am^{-2}}$. It is thereby important to emphasize, that $j_c$ is $10^5$ times smaller than the currents needed to induce a motion in present-day spintorque experiments on ferromagnetic domain walls \cite{grol03,tsoi03,yama04}. However, the rotation occurred only in the presence of a small temperature gradient ($\sim 1\,{\rm K\,cm^{-1}}$) inducing gradients in the relevant forces, which in turn caused rotational torques. It was argued that above the critical current density, $j_c$, the skyrmions start to move and that only the sliding skyrmions can be rotated by the tiny torques. While microscopic probes such as neutron scattering and Lorentz force microscopy may in principle confirm the depinning and motion of the skyrmions, they are not capable of detecting the emergent electrodynamics. Instead, to detect the motion of the skyrmions, measurements of the emergent electric fields are ideally suited, because they are directly proportional to their velocity. This may be readily achieved by means of the Hall effect.

The magnetic properties of MnSi are governed by a combination of ferromagnetic exchange interactions and weak spin-orbit coupling in the absence of inversion symmetry.  At zero magnetic field MnSi displays a paramagnetic to helimagnetic transition at $T_c\approx28.5\,{\rm K}$. In a small range of fields and temperatures below $T_c$ the SLP is stabilized \cite{mueh09}, where the skyrmions (the magnetic whirls) form a hexagonal lattice perpendicular to the applied magnetic field. The lattice constant $2\lambda_{\rm helix}/\sqrt{3}$ of the skyrmion lattice, determined from the reciprocal lattice vectors, is set by the wavelength of the helimagnetic state, $\lambda_{\rm helix}\sim 180\,{\rm \AA}$. 

\begin{figure}[h]
\centerline{\includegraphics[width=9cm,clip=]{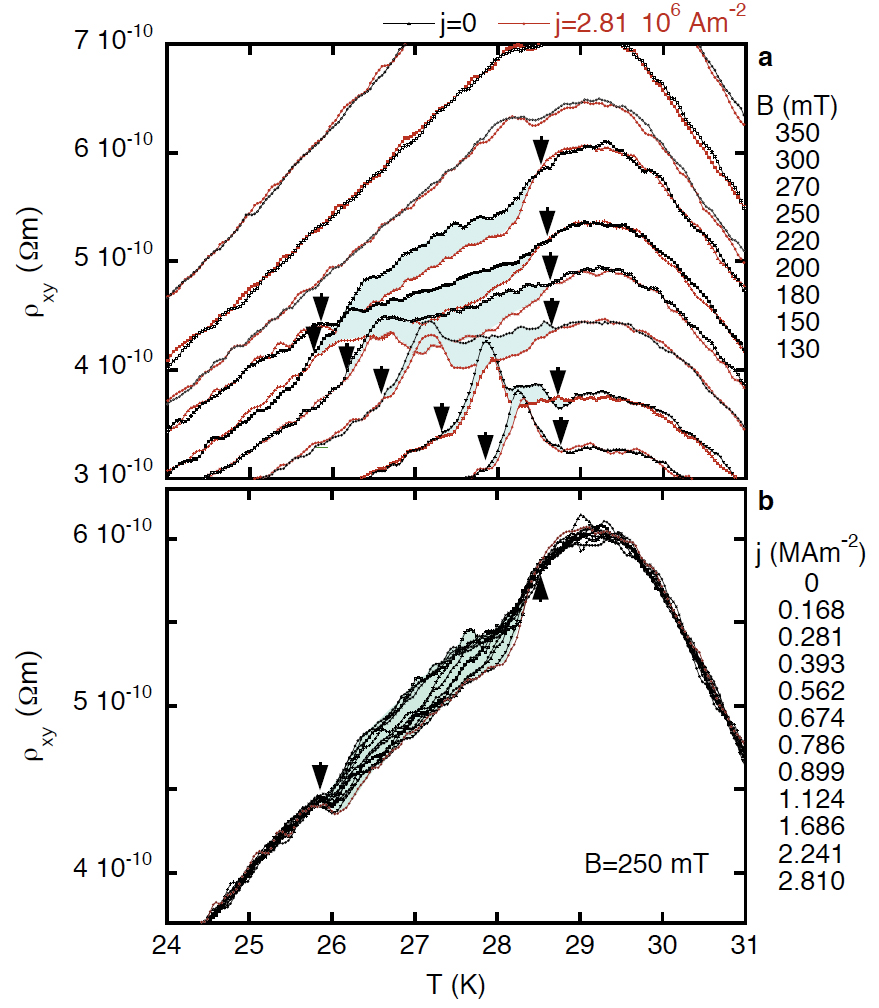}}
\caption{Temperature dependence of the Hall resistivity in the skyrmion lattice phase of MnSi under a large applied DC electric current. To study the effect of the applied DC current it is superimposed on a small AC excitation that allows to detect the signal. (a) Hall resistivity for various magnetic fields. Under an applied DC current of $2.81\cdot10^{6}\,\mathrm{Am^{-2}}$ the Hall signal is suppressed in the entire skyrmion phase (blue shading). (b) Hall resistivity in the SLP for an applied magnetic field $B=250\,\mathrm{ mT}$ under selected applied DC currents between $j=0$ and $j=2.81\cdot10^{6}\, \mathrm{Am^{-2}}$.} 
\label{figure2}
\end{figure}

Our study of the influence of an electric current on the SLP was carried out on high-purity single crystals using a standard six-terminal lock-in technique (see Materials and Methods for details). Shown in Fig.\,\ref{figure2}\,(a) are typical temperature dependences of the Hall resistivity, $\rho_{xy}$, for small currents (black curves). Its dominant features arise from the temperature dependence of the anomalous Hall effect. The small maximum  in $\rho_{xy}$ is thereby a characteristic of the magnetization at the lower boundary of the SLP above a narrow regime of phase coexistence between the conical phase and the SLP \cite{Bauer2010}. Also shown in Fig.\,\ref{figure2}\,(a) is the Hall resistivity under an applied DC current density of $j=2.81\cdot10^{6}\,\mathrm{Am^{-2}}$, where a suppression (marked in light blue shading) is observed in the temperature range marked by the black arrows. The latter is in excellent agreement with the phase boundaries of the SLP as inferred from the magnetization and susceptibility. The size of the suppression of the Hall signal is similar to the topological Hall contribution $\Delta \rho_{xy}\approx 4\,{\rm n\Omega cm}$ previously inferred \cite{neub09} from Hall effect measurements at small currents \cite{supplement}.

\begin{figure}[h]
\centerline{\includegraphics[width=9cm,clip=]{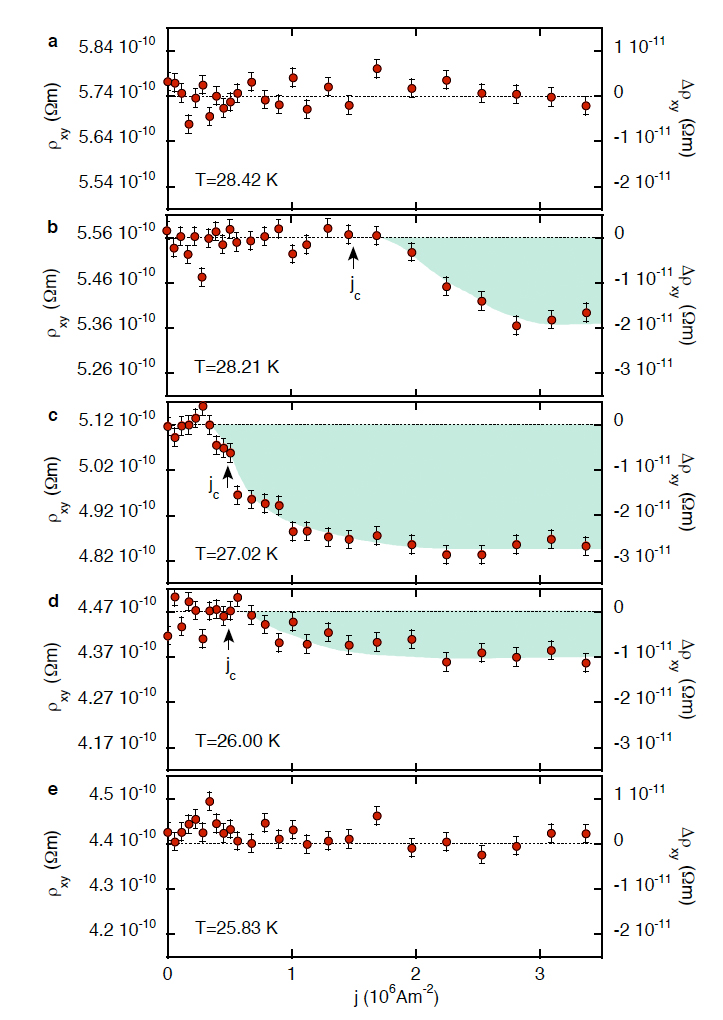}}
\caption{Typical variation of the Hall resistivity, $\rho_{xy}$, of MnSi as a function of applied DC current at $B=250\,\mathrm{mT}$ and for selected temperatures.  For temperatures in the skyrmion lattice phase, shown in panels (b), (c) and (d), the signal is suppressed above an ultralow current density, $j_{c}$, and limits towards a constant value for large currents. Absolute data values are given on the left hand side, while the relative change with respect to $j=0$ is given on the right hand side. 
} \label{figure3}
\end{figure}

Detailed Hall data for an applied magnetic field of 0.25\,T and a wide range of applied DC currents, $j$, are shown in Fig.\,\ref{figure2}\,(b). From these the evolution of $\rho_{xy}$ with applied electric current density was inferred for selected temperatures as shown in Fig.\,\ref{figure3}. At temperatures outside the SLP, shown in panels (a) and (e), the Hall signal is unchanged as a function of $j$. Within the SLP $\rho_{xy}$ is within the experimental accuracy unchanged for small current densities, $j<j_c$, followed by a clear decrease above $j_{c}$  over a finite range of applied currents (light blue shading), which saturates for even larger currents (the method how $j_c$ was determined is described in the supplement \cite{supplement}). 

Shown in Fig.\,\ref{figure4}\,(a) is the critical current density as a function of temperature, where $j_c\sim 10^6\,{\rm A\,m^{-2}}$ agrees within a factor of two with the onset of the rotation of the scattering pattern observed in the SANS study \cite{Jonietz10}. As discussed in Ref.\,\cite{Jonietz10} the exceptionally low value of $j_c$ arises from a combination of several factors. First, the very efficient Berry-phase coupling between conduction electrons and the spin structure.  Second, because of the smooth variation of the magnetisation the skyrmion lattice couples only weakly to defects and the atomic crystal structure. Third, the long range stiffness and crystalline character of the skyrmion lattice \cite{Adams11} leads to a partial cancellation of the pinning forces \cite{schm73,lark74}. $j_c$ grows by roughly a factor of two when approaching the (first order) transition to the weakly field-polarized paramagnetic state. At the same time the suppression of $\rho_{xy}$ given by $\Delta\rho_{xy}^\infty=\Delta \rho_{xy}(j\gg j_c)-\Delta \rho_{xy}(j \ll j_c)$
is largest in the center of the skyrmion phase with a gradual decrease at the lower boundary (cf. Fig.\,\ref{figure4}\,(b)). 

In order to address our observations on $\rho_{xy}$ from a theoretical point of view we note that the forces driving the skyrmion lattice, which also cause the topological contribution to the Hall signal, originate in quantum mechanical phases (Berry phases) picked up by electrons when their spin follows the  orientation $\hat n(\vec r,t)=\vec M/|\vec M|$ of the local magnetization $\vec M$. The Berry phase can be rewritten \cite{volovik87,zhang09,zang11} as an effective Aharonov-Bohm phase associated with `emergent' magnetic and electric fields, $\vec B^e$ and $\vec E^e$. As the Berry phase is given by the solid angle  covered by $\hat n$, the emergent fields $\vec B^e$ and $\vec E^e$ measure the solid angle for an infinitesimal loop in space and space-time, respectively
\begin{eqnarray}\label{beff}
\vec B^e_i=\frac{\hbar}{2} \epsilon_{ijk} \hat n \cdot (\partial_j \hat n \times
\partial_k \hat n), \qquad \vec E^e_i=\hbar\, \hat n \cdot (\partial_i \hat n \times
\partial_t \hat n) \
\end{eqnarray} 
with $\partial_i=\partial/\partial r_i$. Because the sign of the Berry phase depends on the spin orientation, a majority spin with magnetization parallel to $\hat n$ carries the emergent charge $q^e_\downarrow=-1/2$ while a minority spin carries the emergent charge $q^e_\uparrow=+1/2$. For a skyrmion, defined as a magnetic whirl where $\hat n$ winds once around the unit sphere in the plane perpendicular to  $\vec B$ (while $\hat{n}$ is constant in $\vec B$ direction), the total `emergent flux' is given by $\int \vec B^e d\vec \sigma=-4\pi \hbar$. It is hence topologically quantized to one flux quantum $- 2 \pi \hbar/|q^e|$ per skyrmion (the sign accounts for the formation of antiskyrmions in MnSi \cite{mueh09}).  Further, according to Eq.~(\ref{beff}) and in complete analogy to Faraday's law of induction a skyrmion lattice drifting with the velocity $\vec v_d$ must induce an electric field $\vec E^e=-\vec v_d \times \vec B^e$, where $E^e/v_d$ inherits its quantization from $B^e$. This topological quantization of $\vec B^e$ and $\vec E^e$ makes skyrmion lattices in metals an ideal system to study quantitatively the emergent electrodynamics underlying the coupling of charge and magnetism \cite{zang11}.  In MnSi the emergent magnetic field acquires an average strength of 2.5\,{\rm T}, i.e. $B^e \approx 2.5\,{\rm T} |e/q^e|,$ where $e$ is the electron charge. 

The total  force on an electron with momentum $\vec k$ and spin orientation
$\sigma$ is therefore given \cite{zang11} by 
\begin{eqnarray}
\vec F_{\sigma \vec k}=e \vec E+\vec F_H+q^e_\sigma (\vec v_{\sigma \vec k n}-\vec
v_d)\times \vec B^e
\end{eqnarray}
 where   $\vec v_{\sigma \vec k n}$ is the velocity of quasi particles in band $n$, 
 $\vec E$ is the physical electrical field and  $F_H \ll e E$ is the Hall force  from the normal and anomalous Hall effect. Further dissipative drag forces  $\vec F_{\rm diss}$ arising for $\vec v_d\neq 0$ are probably much smaller \cite{supplement}. The extra electric current induced by $-q^e_\sigma \vec v_d \times \vec B^e$  transverse to the electrical current has to be canceled exactly by the change of the electric Hall field  $\Delta E_\perp= \Delta \rho_{yx} j=-\Delta \rho_{xy} j$ with $\Delta \rho_{xy}=\rho_{xy}(j)-\rho_{xy}(0)$. For a current in $x$ and a magnetic field in $z$ direction we find
\begin{eqnarray}
\Delta E_\perp & \approx& -\frac{ \Delta
\sigma_{yx} E}{\sigma_{xx}}=-\frac{\Delta j_\perp}{\sigma_{xx}}=- \tilde P \left|\frac{q^e}{e}\right| \vec
E^e_y= \tilde P \left|\frac{q^e}{e}\right|  (\vec v_d \times \vec B^e)_y \label{Eperp}\\
\tilde P&=&\left|\frac{e}{q^e}\right|\frac{ \langle \! \langle  j, j^e \rangle \! \rangle}{ \langle
\! \langle j, j \rangle \! \rangle}\approx -
\frac{\sum_{n, \vec k,\sigma=\pm 1}  \sigma \tau_{\sigma n}(v_{\sigma \vec
k n}^y)^2 \partial_{\epsilon} f^0_{{n \sigma}}}{\sum_{n, \vec k,\sigma=\pm 1} 
\tau_{\sigma n}  (v_{\sigma \vec k n}^x)^2 \partial_{\epsilon} f^0_{n \sigma}}
\label{p}
\end{eqnarray}
where the dimensionless spin polarization $\tilde P$ can be obtained by calculating
the cross correlation of the charge current $j$ and the emergent current $j^e$ using Kubo formulas. $\tilde P$ is the ratio of  electric currents obtained from $E^e$ and $E$, where a simple approximation for $\tilde P$ is given in Eq.~(\ref{p}) in the relaxation time approximation of a multiband system ($ f^0_{n \sigma}$ is the Fermi distribution for band $n$ with  scattering rate $1/ \tau_{\sigma n}$ and spin-orientation $\sigma$ relative to the local magnetization).  Up to the factor $\tilde P$, the measurement of the Hall field is therefore a direct measurement of the emergent field $E^e$ and of $v_{d}=E^e/B^e$ since $B^e$ is quantized.

The drift velocity $v_d$ in absolute units is  obtained from   Eq. (\ref{Eperp})
\begin{eqnarray}
v_{d\|}&\approx &-\left|\frac{e}{q^e}\right|\, \frac{j \Delta \rho_{xy}}{B^e  \tilde P}=v_{\rm pin} \,\frac{j \Delta \rho_{xy}}{j_c \Delta \rho_{xy}^\infty}\approx  \frac{j}{10^6 \,{\rm Am^{-2}} }  \,\frac{\Delta \rho_{xy}}{\Delta \rho_{xy}^\infty}\,0.12 \frac{\rm mm}{\rm s}.
\label{vdrift}
\\
v_{\rm pin}&\approx&- j_c \left|\frac{e}{q^e}\right|\,   \frac{ \Delta \rho_{xy}^\infty}{B^e \tilde P }
\approx \frac{j_c}{10^6 \,{\rm Am^{-2}} } 0.12 \frac{\rm mm}{\rm s}.\label{absolute}
\end{eqnarray}
Here we used $\Delta \rho_{xy}^\infty\approx -3\cdot10^{-11}\,{\rm\Omega m}$ in the center of the skyrmion phase and estimated the effective polarization to be $\tilde P \approx 0.1$ (cf. Ref.~\cite{neub09}). We further used that  $ \Delta \rho_{xy}^\infty |e|/(B^e \tilde P |q^e|) \approx  -v_s/j=- |q^e/(e M)| \tilde P$ is approximately independent of the local magnetization and therefore of the temperature. As expected, the corresponding pinning velocities (see labels on the right axes in Fig. \ref{figure4}a) are of the same order of magnitude as the electronic drift velocities, $v_{\rm drift} \sim j/en\approx 0.16\,{\rm mm\,s^{-1}}$ for $j\sim j_c \sim 10^6\,{\rm Am^{-2}}$, where we estimate $n\approx 3.8\,\cdot10^{22}\,$cm$^{-3}$ from the normal Hall constant \cite{neub09}. The expression $4 \pi \hbar M v_{\rm pin}$ may finally be interpreted as the force per skyrmion and per length needed to depin the SLP \cite{supplement}.

The critical current and  the pinning velocities grow by a factor of two just below the first order transition $T^c_p$, see Fig. \ref{figure4}\,(a). This cannot be explained by the decrease of the local magnetization (proportional to the small drop in  $|\Delta \rho_{xy}^\infty|$). Instead the increase of $j_c$ reflects most likely the reduction of the stiffness of the skyrmion lattice when approaching $T^c_p$, which is only a very weak first order phase transition. With decreasing stiffness the skyrmion lattice adjusts better to the disorder potential implying that the pinning forces increase substantially  \cite{lark74,blatter94,natter04}. Consistent with this picture, there is essentially no temperature dependence of $j_c$ at the low temperature side of the SLP, where the phase transition is strongly first order. Here $\Delta\rho_{xy}^{\infty}$ shows a gradual temperature dependence in the range from 25.8\,K to 26.5\,K that may be attributed to the phase coexistence between conical phase and skyrmion phase as inferred from high precision magnetometry \cite{Bauer2010}. 

Combining the results of our analysis, we show in Fig.~\ref{figure4}\,(c) a scaling plot of $\Delta E_\perp=-\Delta \rho_{xy} j$, or equivalently of the parallel drift velocity $v_{d\|}$ and the associated emergent electric field $E^e_{\perp}$ as a function of $j/j_c$. To obtain quantitative values for $v_{d\|}$ and $E^e_\perp$ in physical units, one may refer to Fig.~\ref{figure4}\,(a), Eq. (\ref{absolute}), and $E^e_\perp=v_{d\|} B^e$ with $B^e=2.5\,{\rm T} |e/q^e|$.

For $j<j_c$ the drift velocity vanishes within our experimental precision as the skyrmion lattice is pinned by disorder. For $j>j_c$ the Magnus forces are sufficiently strong to overcome the pinning forces and the skyrmions start to move. For $j\gg j_c$ their velocity becomes proportional to the current as pinning forces can be neglected. This is precisely the picture which has been developed for the depinning transition of charge density waves and vortices \cite{schm73,lark74,blatter94,natter04}. However, the precise behavior of skyrmions differs from that of vortices because their dynamics is very different. Due to their strong friction, superconducting vortices flow approximately perpendicular to the current following the Magnus force and the resulting Hall signals are tiny \cite{kopnin02,blatter94}. In contrast, skyrmions drift dominantly parallel to $\vec j$ thus reducing the relative speed of spin current and giving rise to a large Faraday field in the perpendicular direction. A more technical discussion of the validity of the scaling, the forces on the SLP and its direction of motion is given in \cite{supplement}. 

The direct observation of the emergent electric field of skyrmions reported in this paper allowed us to measure their depinning transition and subsequent motion quantitatively. This opens the possibility to address fundamental questions of the coupling of magnetism, electric currents and defects, respectively. The control and detection of the motion of magnetic whirls (skyrmions) by the interplay of emergent and real electrodynamics thereby promises to become an important route towards spintronic applications.

\newpage
\section*{Acknowledgements}

We wish to thank P. B\"oni, H. Hagn, N. Nagaosa, T. Nattermann, S. Mayr, M. Opel, B. Russ, B. Spivak, G. St\"olzl and V. M. Vinokur for helpful discussions and support. RR, AB, MW and CF acknowledge financial support through the TUM Graduate School. KE acknowledges financial support through the Deutsche Telekom Stiftung and the Bonn Cologne Graduate School. Financial support through DFG TRR80, SFB608 and FOR960 is gratefully acknowledged.

\section*{Author Contributions}
TS, RR MH, MW and CP developed the experimental set-up; 
TS and RR performed the experiments; 
TS, RR and CP analysed the experimental data;
CF wrote the software for analysing the data;
AB grew the single-crystal samples and characterised them; 
KE, MG and AR developed the theoretical interpretation; 
CP supervised the experimental work; 
CP and AR proposed this study and wrote the manuscript; 
all authors discussed the data and commented on the manuscript;
correspondence should be addressed to CP and AR.

\newpage


\section*{Methods}

\textbf{Sample preparation:} Single crystals of MnSi were grown by optical float-zoning under ultra-high vacuum compatible conditions \cite{Neubauer:RSI2010}. The specific heat, susceptibility, and resistivity of small pieces taken from this single crystal were in excellent agreement with the literature, where the residual resistivity ratio was of the order 100. The latter indicates good, though not excellent sample purity. The sample quality was the same as for the samples studied in the SANS experiments reported in Ref.\,\cite{Jonietz10}. Samples for the measurements reported here were oriented by Laue x-ray diffraction, cut with a wire saw, and carefully polished to size. Current leads were soldered to the small faces of the sample, while Au wires for the voltage pick-up were spot-welded onto the surface of the sample.

\textbf{Spin torque transport:} 
For our measurements of the Hall and longitudinal resistivity we modified a standard six-terminal phase-sensitive detection system such that large DC currents could be superimposed on a small AC excitation. The set-up is based on a method used for measurements of superconducting tunnel junctions \cite{Welter:2007}. It was tested on high-purity Cu to ensure proper operation. In all experiments the AC excitation amplitude were not larger than a few \% of the applied DC currents. The samples were carefully anchored to the cryogenic system to minimize ohmic heating and temperature gradients. In particular, compared to the SANS experiments reported in Ref.\,\cite{Jonietz10} all thermal gradients were minimized. The Hall signal and the longitudinal resistivity, $\rho_{xy}$ and $\rho_{xx}$, respectively, were measured simultaneously at a low excitation frequency of 22.5\,Hz. To correct for the remaining tiny amount of uniform ohmic heating, which generated a small systematic temperature difference between sample and thermometer of less than a few tenths of a K for the largest currents applied, we calculated from the longitudinal resistivity $\rho_{xx}$ the true sample temperature. However, we have tested carefully that our results do not depend on the precise way how these small ohmic heating effects are corrected. 

\newpage

\clearpage \thispagestyle{empty}

\clearpage \thispagestyle{empty}

\clearpage \thispagestyle{empty}
\begin{figure}[h]
\centerline{\includegraphics[width=9cm,clip=]{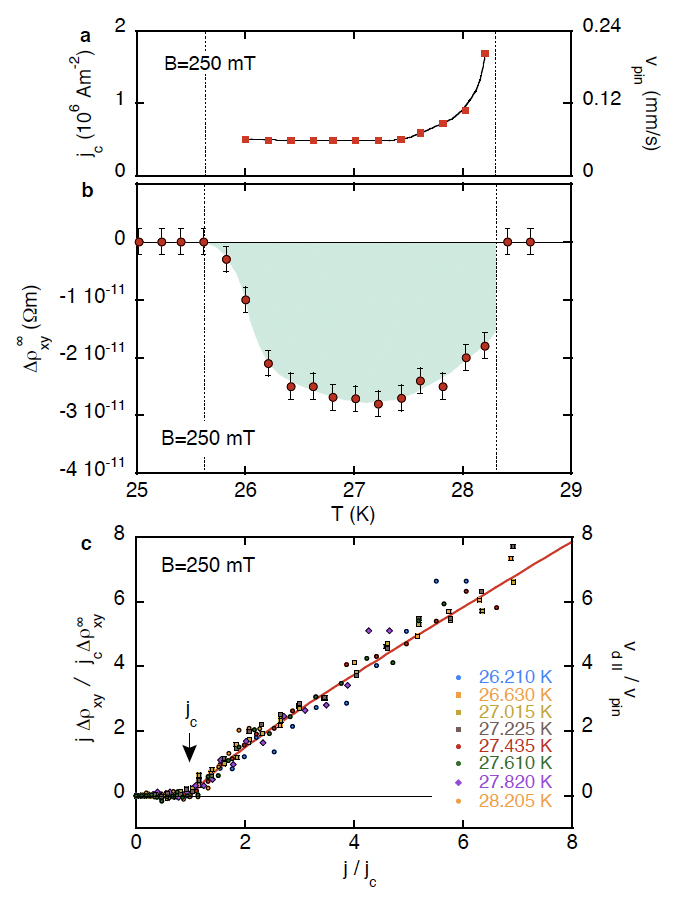}}
\caption{Main characteristics of the response of the skyrmion lattice phase in MnSi under applied electric currents. 
(a) Temperature dependence of the critical current density, $j_c$. The label on the right axis gives an estimate of the pinning velocity, see Eq.~(\ref{absolute}).
(b) Total change of the Hall resistivity  for large currents, $\Delta \rho_{xy}^{\infty}=\rho_{xy}(j\ll j_c)-\rho_{xy}(j\gg j_c)$, as a function of temperature. 
(c) Scaling plot of the transverse electric field, $\Delta E_{\perp}=-j \Delta \rho_{xy}$, in units of  $j_c \Delta \rho_{xy}^\infty$ induced by the moving skyrmion lattice. As $\Delta E_\perp$ is proportional to the emergent field $E^e$, this also constitutes a scaling plot of the emergent electric field $E^e$ (in units of $v_{\rm pin} B^e$) or
of the drift velocity $v_{d\|}$ in units of the pinning velocity $v_{\rm pin}$  (shown in  (a)). Unscaled data used to construct panel (c) are shown in the supplement \cite{supplement}.
} \label{figure4}
\end{figure}


\end{document}


\newcommand{\OLD}[1]{{\tiny {\bf old:} #1 }}
\newcommand{\NEW}[1]{{ \it #1 }}
\renewcommand{\vec}[1]{\boldsymbol{#1}}
\newcommand{\w}{\omega}
\newcommand{\zrzn}{ZrZn$_2$}
\newcommand{\ZrZn}{ZrZn$_2$}
\newcommand{\uge}{UGe$_2$}
\newcommand{\MnSi}{MnSi\,-\,110 }
\newcommand{\Mnsi}{MnSi\,-\,111 }
\newcommand{\Tc}{$T_{c}$ }
\newcommand{\rhoxx}{$\rho_{xx}$ }
\newcommand{\rhoxy}{$\rho_{xy}$ }

\renewcommand{\floatpagefraction}{0.5}
\bibliographystyle{nature}




\title{Supplementary Information for:\\
Emergent electrodynamics of skyrmions in a chiral magnet}



\author{T. Schulz}
\affiliation{Physik Department E21, Technische Universit\"at M\"unchen, D-85748 Garching, Germany}

\author{R. Ritz}
\affiliation{Physik Department E21, Technische Universit\"at M\"unchen, D-85748 Garching, Germany}

\author{A. Bauer}
\affiliation{Physik Department E21, Technische Universit\"at M\"unchen, D-85748 Garching, Germany}

\author{M. Halder}
\affiliation{Physik Department E21, Technische Universit\"at M\"unchen, D-85748 Garching, Germany}

\author{M. Wagner}
\affiliation{Physik Department E21, Technische Universit\"at M\"unchen, D-85748 Garching, Germany}

\author{C. Franz}
\affiliation{Physik Department E21, Technische Universit\"at M\"unchen, D-85748 Garching, Germany}

\author{C. Pfleiderer}
\affiliation{Physik Department E21, Technische Universit\"at M\"unchen, D-85748 Garching, Germany}

\author{K. Everschor}
\affiliation{Institute for Theoretical Physics, Universit\"at zu K\"oln, D-50937 K\"oln, Germany}

\author{M. Garst}
\affiliation{Institute for Theoretical Physics, Universit\"at zu K\"oln, D-50937 K\"oln, Germany}

\author{A. Rosch}
\affiliation{Institute for Theoretical Physics, Universit\"at zu K\"oln, D-50937 K\"oln, Germany}

\date{\today}

\begin{abstract}
Supplementary information for our manuscript, 
entitled 'Emergent electrodynamics of skyrmions in a chiral magnet' is presented. The data are shown in a form essentially as measured, highlighting the determination of the critical current density. A detailed technical discussion is presented of the forces acting on the skyrmion lattice under electric current flow followed by a discussion of the relative size of the Hall effect and the Faraday effect in multi-band and Galilean invariant systems.
\end{abstract}

\maketitle

\newpage
In the following we present at first unscaled data of the main effect reported in the main paper. This is followed by a discussion of the forces on the skyrmion lattice. We then show how  the pinning force and the direction of motion of the skyrmions can be described theoretically and present arguments that the dissipative and drag forces on the electrons are very small. Finally, we will discuss the relative size of the emergent Hall effect and the Faraday effect both in a material with many bands and in a Galilean invariant system.

\section{Drift velocities inferred from experimental data}

\begin{figure}[h]
\centerline{\includegraphics[width=13cm,clip=]{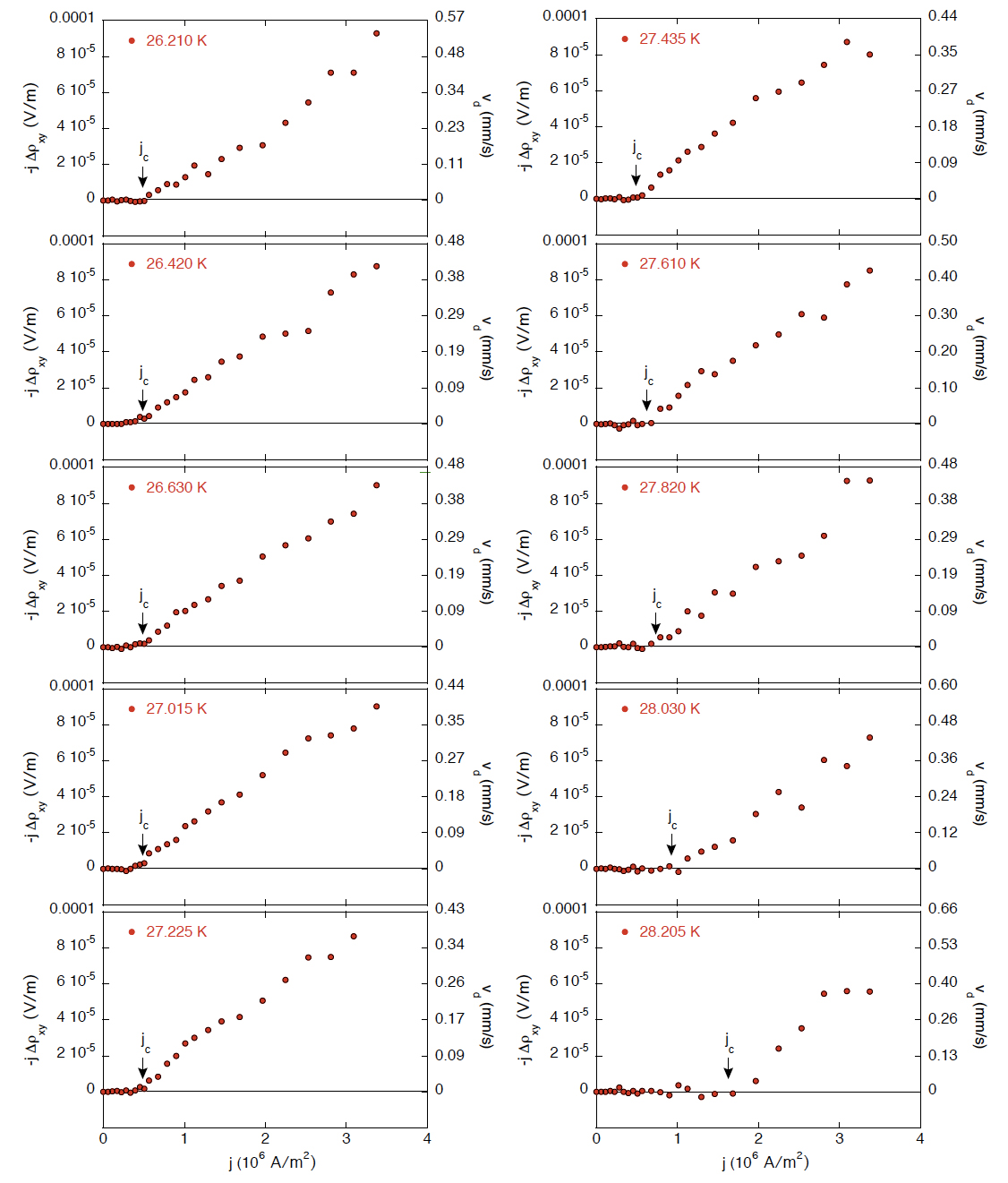}}
\caption{\label{transE} Typical data of the product of the current density $j$ with the change of the Hall resistivity $\Delta\rho_{xy}$ as a function of current density. The departure of the data points from zero marks the critical current density, where the skyrmion lattice begins to move. The scale on the right hand side displays the corresponding drift according to Eq.\,(5) of the main text.} 
\label{figure-S1}
\end{figure}

Shown in Fig.\,\ref{figure-S1} are typical data of the product of the current density $j$ with the change of Hall resistivity, $\Delta\rho_{xy}$, recorded at an applied field of 250\,mT. These form the basis of the scaling shown in Fig.\,3\,(c) of the main text. The data as plotted here was also used to determine the critical current density $j_c$ at the point where $j\,\Delta\rho_{xy}$ begins to deviate from zero.  Shown on the right hand side of each panel is the corresponding drift velocity according to Eq.\,(5) of the main text. 

\section{Forces on the skyrmion lattice and depinning transition}
To understand the origin of the drift velocity $\vec v_d$ of the skyrmions, we study the relevant forces by projecting the Landau-Lifshitz-Gilbert equation \cite{zhan04} on the translational modes using an approach pioneered by Thiele \cite{Thiele72}. Besides a Magnus force \cite{Jonietz10}, which is the counterforce to the emergent Lorentz force arising from $\vec B^e$, and dissipative forces one has also to consider pinning forces $\vec f_{\rm pin}$ arising from  inhomogeneities which prohibit the motion of skyrmions for small currents. Balancing all forces (per skyrmion, per length, and per $\hbar M$)  one obtains
\begin{eqnarray}\label{Sforces}
\vec G &\times& (\vec v_s-\vec v_d)+ {\mathcal D} (\beta \vec v_s - \alpha \vec
v_d)={\vec f}_{\rm pin}\\
{\vec f}_{\rm pin}&\approx& -4 \pi v_{\rm pin} f(v_d/v_{\rm pin}) \hat v_d
\end{eqnarray}
where $\alpha$ and $\beta$ parametrize the damping \cite{zhan04}, the gyromagnetic coupling vector $\vec G=\int_{\rm uc} \vec B^e/\hbar$ integrated over the 2d magnetic unit cell evaluates to $-4 \pi \hat{\vec B}$, $\vec v_s=\langle \vec j^e \rangle/M$ is the velocity of the current associated to the charges $q^e_{\sigma}$, and  ${\mathcal D}=\int_{\rm uc} (\partial \hat n)^2/2$ \cite{Thiele72}. Up to tiny corrections  of the order $\sim \sigma_{xy}/\sigma_{xx}$, $\vec v_s$ is  parallel to the applied current. For the skyrmions in MnSi, the dissipative terms  are expected to be small as $\alpha$, $\beta \ll 1$ arise mainly from spin-orbit effects and spin flip scattering (see Sec. \ref{supp2} below) and  we obtain values of order 1 for  $\mathcal D/|G|$ for typical parameters \cite{mueh09} where the skyrmion crystal is stable. In the following, we therefore neglect the dissipative forces in the parameter range of our experiment.

The small size of the dissipative forces has important consequences for the motion of the skyrmions, especially in the regime $j \gg j_c$ where the pinning force $\vec f_{\rm pin}$ (discussed below) can be neglected. In this regime, one obtains $\vec v_d \approx \vec v_s$ with corrections of order $\alpha D/|G|$, i.e., the skyrmions move {\it parallel} to the current in such a way that the Magnus force is canceled. In contrast, for vortices in superconductors, where $\alpha D/|G| \gg 1$ \cite{kopnin02}, dissipation dominates such that $v_d \ll v_s$. Therefore the motion of superconducting vortices is dominantly perpendicular to the current following the direction of the Magnus force. Similar to the skyrmions, charge density waves \cite{natter04} flow parallel to the current. However, this is less suprising since the dominant force on CDWs is not a Magnus force but arises directly from the electric field.

The pinning force, $\vec f_{\rm pin}$, for the moving skyrmion lattice is oriented antiparallel to the drift velocity $\vec v_d$. We parametrize its strength  by the 'pinning velocity' $v_{\rm pin}$ and a dimensionless function $f(v_d/v_{\rm pin})$ with $f(0)=1$ and  $f(x \gg 1)\sim x^\alpha$, $\alpha<1$ (assuming the absence of creep). $f$ is not known, but we expect a behavior similar to that of superconducting vortices though not identical due to the different dynamics \cite{schm73,lark74,nattermann92}. Furthermore, we do not expect that $f$ is a fully universal function; it will depend, e.g., on the various elastic constants of the skyrmion lattice. Nevertheless, the observed approximate scaling of $v_d/v_{\rm pin}$ as a function of $v_s/v_{\rm pin}\approx j/j_c$ shown in Fig.~3\,(c) of the main article suggests that $f$ depends only very little on the distance to the transition temperature while $v_{\rm pin}$ varies strongly.

Neglecting damping, one can calculate $\vec v_d/v_{\rm pin}$ from $f$. For unkown $f$, one may determine $f$ from the measurement of $v_{d \|}/v_{\rm pin}$ as a function of $j/j_c$. By projecting Eq.~(\ref{Sforces}) (for $\alpha,\beta=0$) to the direction perpendicular and parallel to $\vec v_d$ and using $v_d^2=v_{d\|}^2+v_{d\perp}^2$ and Eq. (5) of the main article, one obtains for the components of $\vec v_d$ parallel and perpendicular to $\vec v_s$ the following equations
\begin{eqnarray}
\frac{v_{d\|}}{v_{\rm pin}}&\approx & \frac{j}{j_c} \frac{\Delta \rho_{xy}}{\Delta \rho_{xy}^\infty}\label{vdp} \\
\frac{v_d}{v_{\rm pin}}&\approx &\sqrt{\frac{v_{d\|} v_s}{v_{\rm pin}^2}}\approx \frac{j}{j_c} \sqrt{\frac{\Delta \rho_{xy}}{\Delta \rho_{xy}^\infty}} \label{vdE} \\
\frac{v_{d\perp}}{v_{\rm pin}}&\approx& \frac{1}{v_{\rm pin}} \sqrt{v_{d\|}(v_s-v_{d\|})}\approx \frac{j}{j_c} \sqrt{\frac{\Delta \rho_{xy}}{\Delta \rho_{xy}^\infty}\left(1-\frac{\Delta \rho_{xy}}{\Delta \rho_{xy}^\infty}\right)}\\
\theta&=&\arccos \frac{v_{d\|}}{v_{d}}\approx \arccos \sqrt{\frac{\Delta \rho_{xy}}{\Delta \rho_{xy}^\infty}} \\
f&\approx & \frac{v_s v_{d\perp}}{v_d v_{\rm pin}}\approx \frac{v_s  \sqrt{v_{d\|}(v_s-v_{d\|})}}{v_{\rm pin} \sqrt{v_{d\|} v_s}}\approx \frac{j}{j_c} \sqrt{1-\frac{\Delta \rho_{xy}}{\Delta \rho_{xy}^\infty}} \label{fE}
\end{eqnarray}
While Eq.~(\ref{vdp}) (which just copies Eq. (5) of the main article) reflects that the Hall effect provides a direct measurement of the drift velocity of skyrmions in parallel direction, the other equations show, that -- under the assumption that  Eq. (\ref{Sforces}) is valid and that damping is negligible --  one can also obtain the drift velocity in perpendicular direction, the angle $\theta$ between $\vec v_d$ and the current and the pinning force as function of $v_d/v_{\rm pin}$ experimentally.

\begin{figure}[h]
\centerline{\includegraphics[width=10cm,clip=]{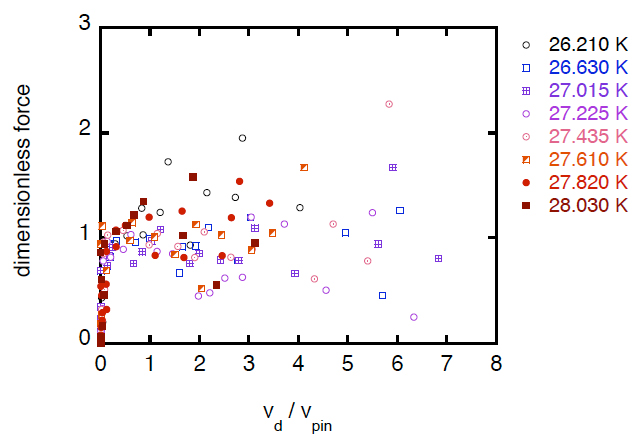}}
\caption{\label{figForce} Force $f$ on the skyrmions in dimensionless units as a function of $v_d/v_{\rm pin}$ from Eqs. (\ref{vdE}) and (\ref{fE}).} 
\label{figure-S2}
\end{figure}

Unfortunately, our present data is too noisy to extract reliably the force $f$ or $v_{d \perp}$. As an example, we show in Fig.~\ref{figForce} the dimensionless force $f$ as a function of $v_d/v_{\rm pin}$ obtained from Eqs. (\ref{vdE}) and (\ref{fE}).  For $j<j_c$ the force grows with the applied current but $v_d$ remains approximately $0$ which results in a vertical line for $f<1$, $v_d=0$. For $v_d>0$, the force appears to remain approximately constant as a function of $v_d$, but due to the large noise a definite conclusion is presently not possible. 

\section{Dissipative drag forces and corrections to the adiabatic approximations} \label{supp2}

In this section we present arguments why dissipative drag forces acting on the electrons may be neglected. Since we have argued that dissipative forces acting on the skyrmion lattice are small this is not surprising. In addition we briefly discuss in this section the role of other corrections which may modify the emergent magnetic and electric fields.

When the skyrmion lattice starts to move, the electrons experience the forces  induced by the emergent magnetic and electric fields $\vec B^e$ and $\vec E^e$, discussed in the main text. In addition dissipative forces $\vec F_{d}$ drag the electrons parallel to the motion of the skyrmion lattice as the skyrmion lattice induces effectively also a weak periodic potential arising from at least three different sources, notably spin-orbit coupling in the band structure, small variations of the amplitude of the magnetization, and variations of $\vec B^e$ around its average value. 

To obtain an estimate of the strength of these forces for small drift velocities $\vec v_d$ of the skyrmion lattice, it is useful to make use of results for the generic problem of a weak time-dependent periodic electric potential $\Phi(\vec r - \vec v_d t)$ moving in a diffusive system described by the diffusion equation for the charge density $n$
\begin{eqnarray}
\partial_t n=\vec \nabla (D(n) \vec \nabla n)+\vec \nabla (\sigma(n) \vec \nabla \Phi(\vec r - \vec v_d t))
\end{eqnarray}
In a comoving frame of reference one obtains  a stationary solution by the transformation $n \to n(\vec r - \vec v_d t)$ and $\vec r \to \vec r+ \vec v_d t$ with
\begin{eqnarray}
- \vec v_d \vec  \nabla n=\vec \nabla D(n) \vec \nabla n+\vec \nabla \sigma(n) \vec \nabla \Phi(\vec r)
\end{eqnarray}
with the density-dependent diffusion constant $D(n)$ and conductivity $\sigma(n)$.
Integrating this equation with an integration constant $-\vec v_d n_0$ and multiplying with $\vec v_d$ gives
\begin{eqnarray}
-  v_d^2 (n-n_0)= D(n) \vec v_d \vec \nabla n+\sigma(n) \vec v_d \vec \nabla \Phi(\vec r)
\end{eqnarray}
Linearizing the equation using the approximation $D(n)=D={\rm const.}$ and $\sigma(n)=\sigma={\rm const.}$ one finds after Fourier transformation with $n(\vec r)=\sum_{\vec q} n_{\vec q} e^{ i \vec q \vec r}$ 
\begin{eqnarray}
n_{\vec Q_i}\approx  \frac{- i  (\vec v_d \vec Q_i) \sigma \Phi_{\vec Q_i}}{v_d^2+  i   (\vec  Q_i \vec v_d) D}
\end{eqnarray}
where $\vec Q_i\neq 0$ are the reciprocal lattice vectors of the periodic potential. To this linear order no dc current is obtained, while only a small periodic modulation of the density is induced by the potential. To obtain a uniform electric current, one has to expand one order higher in $\Phi$.  Expanding $D$ and $\sigma$ to linear order in the changes of $n$ gives rise to an average current in the direction of $\vec v_d$
\begin{eqnarray}
 j&=&- 1/(V v_d) \int (D(n) \vec v_d \vec \nabla n+\sigma(n) \vec v_d \vec \nabla \Phi(\vec r))\\
&\approx& v_d \sigma \frac{\partial \sigma}{\partial n} \sum_i  \frac{  (\vec v_d \vec Q_i)^2  |\Phi_{\vec Q_i}|^2}{v_d^4+   ((\vec  Q_i \vec v_d) D)^2}\approx  v_d  \sigma \frac{\partial \sigma}{\partial n} \sum_{\vec v_d \vec Q_i\neq 0}  \frac{   |\Phi_{\vec Q_i}|^2}{ D^2}
\end{eqnarray}
for small $v_d$.
For a potential characterized by variations of order $\Delta V \sim   |e \Phi_{\vec Q_i}|$ and with $\sigma\sim n e \tau/m$, $D\sim v_F^2 \tau$, scattering rate $1/\tau$, mean free path $l$, quasiparticle mass $m$, Fermi energy $\epsilon_F$, Fermi wave length $\lambda_F$, Fermi momentum $k_F$, and Fermi velocity $v_F$,
we can therefore estimate the current and the associated dissipative force from
\begin{eqnarray}
 F_d=\frac{e j}{\sigma} &\sim&   \frac{m v_d}{\tau} \left(\frac{\Delta V}{\epsilon_F}\right)^2
\end{eqnarray}
Note that the current is independent of the scattering rate within our approximation, while the dissipative force is proportional to the scattering rate.

As discussed above, several mechanisms contribute to $\Delta V$ but for a crude estimate of its size it is sufficient to note
that the change of electronic energy due to $\Delta V$, which is of order $(\Delta V)^2/\epsilon_F$ (or much larger), is maximally of the same order as the energy difference of the ferromagnetic and the skyrmion state, $\Delta E_s\sim E_M/(k_F D)^2$, where $E_M$ is a magnetic energy scale and $D$ is the radius of skyrmions
 which for MnSi also describes the distance of skyrmions \cite{neub09}. Therefore we estimate
\begin{eqnarray}
 F_d \lesssim v_d \frac{\hbar}{D^2} \frac{1}{k_F l} \frac{E_M}{\epsilon_F}\sim  v_d B^e \frac{1}{k_F l} \frac{E_M}{\epsilon_F}
\end{eqnarray}
For the last estimate, we used that the emergent magnetic field (quantized to one flux quantum per magnetic unit cell) is of order of
$\hbar/D^2$.

Compared to the force from the emergent electric field, $\vec E^e=-\vec v_d \times \vec B^e$, $F_d$ is strongly suppressed by two small factors, $1/(k_F l)$ and $\frac{E_M}{\epsilon_F}$. Therefore we expect that the contribution of dissipative forces to the topological Hall signal are completely negligible compared to the contributions from $\vec E^e$. Only in a tiny regime close to the depinning transition, well below our experimental resolution, these contributions might have a chance of becoming relevant. In this regime $\vec v_d$ is approximately perpendicular to the current and therefore the contribution of $\vec E^e$ to the Hall effect is suppressed.

Finally, further corrections to the simple formulas for the effective electric and magnetic fields, Eq.(1) in the main text, have to be considered. The formulas for the emergent electric and magnetic fields have been derived under the assumption that the conduction electron spins follow the local magnetization adiabatically and that one does not have to take into account a modification of the bandstructure due to the skyrmion lattice. Due to the smoothness of the skyrmion structure and the large distance of skyrmions of the order of 200\AA, violation of adiabaticity and such bandstructure effects are probably small suggesting that the most important corrections may arise from spin-flip scattering processes which scatter electrons, e.g., from a majority to a minority band (for a quantitative discussion of corrections to adiabatic transport see, e.g., Ref.~\cite{Stiles06}). These spin-flip processes, which can be interaction- or disorder-induced, arise only due to weak spin-orbit scattering. While the non-spin-flip scattering length is estimated to be between 10 and 100\,{\AA}, the spin-flip scattering length is much larger and therefore probably also much larger than the distance between the skyrmions $\sim 200$\,{\AA}.\\

\subsection*{Relative size of emergent Hall- and Faraday effect}

Within the experimental errors, we have found that for rapidly moving skyrmions the contribution to the Hall effect arising from the emergent electric field almost cancels the contribution from the emergent magnetic field. In this section we argue, that, generically, the two different contributions are not identical, while there is an exact cancellation in a Galilean invariant system.

In a Galilean invariant system, a state with finite current is obtained by a Galilei transformation to a moving coordinate system where skyrmions and electrons drift with the same velocity and where therefore the force arising from the emergent magnetic field is exactly canceled by the force $-q^e_{\sigma} (\vec v_d \times \vec B^e)$ from the emergent electric field.

Such a cancellation will not occur in a non-Galilean invariant system and several modifications have to be considered. First,  the damping constants $\alpha$ and $\beta$ in Eq.~(\ref{Sforces}) are generically finite and not equal such that the spin velocity $\vec v_s$, which describes the drift of the emergent current $\vec j^e$,  differs from the drift velocity of the skyrmions, $\vec v_d$, even in the absence of pinning. We have, however, argued that damping effects are relatively small, therefore this, very likely, will be a small effect and we expect  $\vec v_s \approx \vec v_d$ in a regime where pinning can be neglected. As a consequence the average force,
$\langle q^e_\sigma (\vec v_{\sigma \vec k n}-\vec v_s) \times \vec{B}^e \rangle$, cancels exactly, but in a complex multiband system (with quasiparticle velocity $v_{\sigma \vec k n}$) it depends on the orientation $\sigma$ of the spin relative to the local magnetization, the momentum $\vec k$, and the bandindex $n$) one nevertheless expects a finite contribution to the Hall effect.

For example, this may already be seen in a simple relaxation time approximation to the Boltzmann equation assuming for simplicity $\vec k$-independent scattering rates, $1/\tau_{\sigma n}$, which depend, however, on the local spin orientation and the band index $n$. For simplicity, we assume a cubic bandstructure without spin-orbit coupling. Within this approximation one obtains for $\sigma_{xy}$
 two contributions arising from $B^e$ and from the emergent electric field for large currents where  $\vec v_d \approx \vec v_s$, describing the emergent Hall and the emergent Faraday effect
\begin{eqnarray}\label{sigF}
\sigma_{xy}^{B^e}&\approx& B^e \sum_{n\sigma} \int \frac{\tau_{\sigma n}^2 q^e_\sigma (v^y_{\sigma \vec k n})^2}{m_{\sigma \vec k n}} \left(-\frac{\partial f^0_{\sigma \vec k n}}{\partial \epsilon}\right)\frac{\mathrm{d}^3 \bf k}{(2 \pi)^3}\\
\sigma_{xy}^{E^e}&\approx&
-\left[ \sum_{n\sigma} \int \tau_{\sigma n} q^e_\sigma (v^y_{\sigma \vec k n})^2 \left(-\frac{\partial f^0_{\sigma \vec k n}}{\partial \epsilon}\right)\frac{\mathrm{d}^3 \bf k}{(2 \pi)^3}\right] B^e
\frac{ \sum_{n\sigma} \int \frac{\tau_{\sigma n} q^e_\sigma}{m_{\sigma \vec k n}}   f^0_{\sigma \vec k n} \frac{\mathrm{d}^3 \bf k}{(2 \pi)^3}}{ \sum_{n\sigma} \int  q^e_\sigma   f^0_{\sigma \vec k n} \frac{\mathrm{d}^3 \bf k}{(2 \pi)^3}}
\end{eqnarray}
where the last fraction describes $v_s/E=j^e/(E M)$. While both expressions contain certain averages over scattering rates, velocities, and the inverse mass $1/m_{\sigma \vec k n}$, in a complex multiband system there is no simple relation between $\sigma_{xy}^{B^e}$ and $\sigma_{xy}^{E^e}$. Localized spins, for example, contribute only to the term in the denominator of $\sigma_{xy}^{E^e}$, thereby reducing the Faraday field. The  special case of localized spins and one fully spin-polarized band was recently discussed in Ref.~\cite{zang11}. But band structure effects can also lead to an enhancement of the emergent Faraday effect relative to the Hall effect when $M$, averaged over all bands, is small. Only in a Galilei invarant single-band model with a constant mass the two contributions cancel exactly, $\sigma_{xy}^{B^e}+
\sigma_{xy}^{E^e}=0$. 
 
It is finally interesting to note that under pressure an apparent increase of the topological contribution of $B^e$ to $\rho_{xy}$ in MnSi by one order of magnitude has been reported \cite{Lee09}. It is presently unclear whether this can be explained by Eq.~(\ref{sigF}) from a change of the scattering rates and/or the band structure with pressure.

\newpage